\documentclass[%
 reprint,
 amsmath,amssymb,
 aps,
 pra,
]{revtex4-1}
\usepackage{subfigure}
\usepackage{float}
\usepackage{hyperref}
\usepackage{cleveref}
\usepackage{nameref}
\usepackage{graphicx}
\usepackage{dcolumn}
\usepackage{bm}
\usepackage{color}
\usepackage{ulem}

\begin{document}

\preprint{APS/123-QED}

\title{Optimized four-qubit quantum error correcting code for amplitude damping channel}

\affiliation{%
 Pritzker School of Molecular Engineering, University of Chicago.
}%

\date{\today}
\newcommand{\chicago}{\affiliation{Pritzker School of Molecular Engineering, University of Chicago, Chicago, Illinois 60637, USA}}

\author{Xuanhui Mao} \chicago
\author{Qian Xu} \chicago
\author{Liang Jiang} \chicago

\begin{abstract}
Quantum error correction (QEC) is essential for reliable quantum information processing. 
Targeting a particular error channel, both the encoding and the recovery channel can be optimized through a
biconvex optimization to give a high-performance, noise-adapted QEC scheme. We solve the biconvex optimization
by the technique of alternating semi-definite programming and identify a new four-qubit code for amplitude damping
channel, one major noise in superconducting circuits and a good model for spontaneous emission and energy dissipation.
We also construct analytical encoding and recovery channels that are close to the numerically optimized ones.
We show that the new code notably outperforms the Leung-Nielsen-Chuang-Yamamoto four-qubit code in terms of the
entanglement fidelity over an amplitude damping channel.
\end{abstract}

\maketitle


\section{\label{sec:level1}Introduction}
Quantum error correction (QEC) \cite{BennettEtAl1996, EkertMacchiavello1996, KnillLaflamme1996, NielsenEtAl1996} addresses errors during quantum computing implementation. Traditional codes are often designed based on the perfect error correction criteria to correct Pauli error in the general Galois field GF(4) \cite{CalderbankEtAl1996} classification.

However, physical systems are often subject to non-Pauli errors, e.g. the amplitude damping errors \cite{Schumacher1996}.  Recognizing the limitations in addressing non-Pauli errors, Leung, Nielsen, Chuang, and Yamamoto introduced an approximate QEC criterion \cite{Leung1997} that tolerents the subtle non-orthogonalities between quantum error syndromes, expanding the scope of QEC codes and enabling the development of codes with shorter block lengths.

Using this approximate criterion and considering the specific structure of noise channels, Leung et al. \cite{Leung1997} developed the first four-qubit quantum error-correcting code designed for amplitude damping errors. Such errors, which describe energy dissipation, are prevalent in quantum systems and differ from traditional Pauli errors. More recently, Dutta et al. have proposed a three-qubit code with post-selection scheme for amplitude damping errors \cite{DuttaMandayam2025}.

While the fidelity of the Leung et al. code marks a significant advancement, it presents opportunities for further improvement. Convex optimization can be employed to derive an optimized numerical recovery matrix \cite{Fletcher2007}. Building upon these developments, this paper employs biconvex optimization \cite{Noh2018,KosutLidar2009}, to simultaneously optimize both encoding and recovery channels for amplitude damping noise. The result is a new optimized four-qubit quantum error-correcting code that surpasses the Leung et al. code in terms of entanglement fidelity. Additionally, we have developed analytical recovery operators for the optimized code that outperforms the original analytical recovery operators acted on the code proposed by Leung et al.

The organization of this paper is as follows: Section II provides a background on optimized quantum error correction. It introduces fundamental definitions and reviews previous research relevant to optimal Quantum Error Recovery (QER) and biconvex optimization. Section III is divided into three subsections. Section III A introduces our new optimized four-qubit amplitude damping code. Section III B discusses the approximate QEC criteria, detailing how our code conforms to these standards. Finally, Section III C elaborates on both the analytical and numerical recovery channels, and it compares the performance of our optimized code with the Leung et al. four-qubit amplitude damping code.

\section{Background}
This section introduces the basics of QEC and the amplitude damping channel, along with a summary of prior work.

\subsection{Basic Definitions}
\subsubsection{QEC Scheme}
Consider a $d$-dimensional Hilbert space $\mathcal{H}_0$. In this space, a QEC scheme can be established as follows: Initially, a state $|\Phi \rangle$ is prepared in a noiseless local memory. This state is encoded from $\mathcal{H}_0$ into a larger Hilbert space $\mathcal{H}_n$ (where $n \geq d$) to introduce redundancy. This encoding is performed via a Completely Positive and Trace-Preserving (CPTP) map $\mathcal{E}: \mathcal{L}(\mathcal{H}_0) \rightarrow \mathcal{L}(\mathcal{H}_n)$. The encoded state traverses a noisy channel $\mathcal{N}: \mathcal{L}(\mathcal{H}_n) \rightarrow \mathcal{L}(\mathcal{H}_n)$. The receiver applies a recovery map $\mathcal{R}: \mathcal{L}(\mathcal{H}_n) \rightarrow \mathcal{L}(\mathcal{H}_n)$ to correct errors introduced by $\mathcal{N}$, followed by a decoding map $\mathcal{D}: \mathcal{L}(\mathcal{H}_n) \rightarrow \mathcal{L}(\mathcal{H}_0)$, restoring the state to $\mathcal{H}_0$. The final state obtained is
\begin{equation}
\hat{\rho} = (\mathcal{D \cdot R \cdot N \cdot E})(|\Phi\rangle \langle\Phi|).
\end{equation}
\\
Fidelity quantifies the similarity between input and output states in QEC. Our paper uses entanglement fidelity to quantify the effectiveness of a QEC scheme in preserving entanglement. High entanglement fidelity indicates robust preservation of quantum correlations against channel-induced errors. To elucidate, consider a maximally entangled state prepared in a noiseless local memory:
\begin{equation}
    |\Phi^+ \rangle \equiv \frac{1}{\sqrt{d}}\sum_{i=0}^{d-1}|i_{\mathcal{H}_0}\rangle |i_{\mathcal{H'}_0}\rangle. 
\end{equation} 

Let $\mathcal{A}$ represent the channel through which the state passes. The entanglement fidelity $F_{ent}(\rho, \mathcal{A})$ \cite{ReimpellWerner2005} is then given by:
\begin{equation}
F_{ent}(\rho, \mathcal{A}) = \langle\Phi^+|(\mathcal{I}_{\mathcal{H}_{0}}\otimes \mathcal{A})(|\Phi^+\rangle \langle \Phi^+|)|\Phi^+\rangle.
\label{eq:fid}
\end{equation}
    
\subsubsection{Choi Matrix}
For a linear map $\mathcal{A}$ that operates from $\mathcal{L}(\mathcal{H}_1)$ to $\mathcal{L}(\mathcal{H}_2)$, the Choi matrix $\hat{X}_{\mathcal{A}}$ is an operator in $\mathcal{L}(\mathcal{H}_1) \otimes \mathcal{L}(\mathcal{H}_2)$, defined by:
\begin{equation}
    (\hat{X}_{\mathcal{A}})_{[ij],[i'j']} = \langle j_{\mathcal{H}_2}|\mathcal{A}(|i_{\mathcal{H}_1}\rangle 
    \langle i'_{\mathcal{H}_1}|)|j'_{\mathcal{H}_2}\rangle.
    \end{equation}
Each element of $\hat{X}_{\mathcal{A}}$ corresponds to the action of $\mathcal{A}$ on a specific basis element of $\mathcal{H}_1$ and $\mathcal{H}_2$, where $|i_{\mathcal{H}_1}\rangle$, $|i'_{\mathcal{H}_1}\rangle$, $|j_{\mathcal{H}_2}\rangle$, and $|j'_{\mathcal{H}_2}\rangle$ form the orthonormal bases of $\mathcal{H}_1$ and $\mathcal{H}_2$, respectively.

According to Choi's theorem \cite{Choi1975}, $\mathcal{A}$ is completely positive if and only if $\hat{X}_{\mathcal{A}}$ is a positive semi-definite matrix. Furthermore, $\mathcal{A}$ is trace-preserving if:
\begin{equation}
\operatorname{Tr}_{\mathcal{H}_2}\hat{X}_{\mathcal{A}} = \hat{I}_{\mathcal{H}_1}.
\label{eq:cptp}
\end{equation}

\subsubsection{Amplitude Damping Channel}
The amplitude damping channel describes the process of energy loss in a quantum system. It is represented by the operators:
\begin{equation}
    A_0=\begin{bmatrix}
    1 & 0 \\
    0 & \sqrt{1-\gamma} 
    \end{bmatrix}    \hspace{1cm}     A_1=\begin{bmatrix}
    0 & \sqrt{\gamma} \\
    0 & 0 
    \end{bmatrix}
\end{equation}
where $\gamma$ denotes the probability of transition from the state $|1\rangle$ to $|0\rangle$, symbolizing the decay process.

In the context of a four-qubit system, the amplitude damping operator $\mathcal{E}^{(n)}_{\tilde{k}}$ is defined as a tensor product of individual qubit operators:
\begin{equation}
\mathcal{E}^{(n)}_{\tilde{k}} = A_{k_1} \otimes A_{k_2} \otimes A_{k_3} \otimes A_{k_4}.
\end{equation}
The superscript $(n)$ refers to the order of amplitude damping error. And $\tilde{k} = {k_1k_2k_3k_4}$, where each $k_i \in {0,1}$ for $i \in {1,2,3,4}$, representing the error indices for each qubit.

The overall quantum noise channel for the four-qubit system undergoing the amplitude damping noise channel, is described by:
\begin{equation}
\mathcal{E}(\rho) = \sum_{\tilde{k}} \mathcal{E}^{(n)}_{\tilde{k}} \rho \mathcal{E}^{\dagger(n)}_{\tilde{k}}.
\end{equation}

\subsection{Biconvex Optimization}

Fletcher, et al. \cite{Fletcher2007} employ convex optimization to find the optimal recovery channel for a fixed encoding, aiming to maximize entanglement fidelity. Noh et al.\cite{Noh2018} and Kosut et al.\cite{KosutLidar2009} further advance this approach by employing biconvex optimization, which optimizes both the encoding and the recovery channel simultaneously to achieve maximal entanglement fidelity.

Entanglement fidelity, denoted as $F_{\text{ent}}(\hat{\rho})$, defined in Eq.~\ref{eq:fid}, is calculated as follows:
\begin{equation}
F_{\text{ent}}(\hat{\rho}) = \frac{1}{d^2} \operatorname{Tr}[\hat{X}_{\mathcal{D}\cdot \mathcal{R}} f_{\mathcal{N}}(\hat{X}_{\mathcal{E}})],
\end{equation}
where $f_{\mathcal{N}}$ represents a linear map defined by:
\begin{equation}
(f_{\mathcal{N}}(\hat{X}))_{l'i',li} = \sum_{k,k' = 0}^{n-1} (\hat{X}_{\mathcal{N}})_{kl,k'l'} (\hat{X})_{ik,i'k'},
\end{equation}
for all $l,l' \in \{0, \ldots, n-1\}$. Here, $\hat{X}_{\mathcal{D}\cdot \mathcal{R}}$ and $\hat{X}_{\mathcal{E}}$ represent the Choi matrices of the composite decoding and recovery channels $\mathcal{D}\cdot \mathcal{R}$ and the encoding channel $\mathcal{E}$, respectively. In the optimization approach, the recovery and decoding channel are treated as a single entity. The bilinear relationship between $F_{\text{ent}}$, $\hat{X}_{\mathcal{E}}$, and $\hat{X}_{\mathcal{D}\cdot \mathcal{R}}$ indicates that simultaneous optimization of the encoding and recovery processes is possible, allowing for a better entanglement fidelity.

Given that both encoding and recovery operators are required to satisfy the CPTP criteria, the biconvex optimization problem can be formulated as:
    \begin{align}
    F_{ent} = \underset{\hat{X}_\mathcal{E},\hat{X}_\mathcal{D\cdot R}}{max}\,\operatorname{Tr}[\hat{X}_\mathcal{D\cdot R}f_{\mathcal{N}}(\hat{X}_\mathcal{E})] \\
    s.t. \,\,\hat{X}_\mathcal{D\cdot R} = \hat{X}_\mathcal{D\cdot R}^\dagger \succeq 0, \,\operatorname{Tr}_{\mathcal{H}_0}\hat{X}_\mathcal{D\cdot R} = \hat{I}_{\mathcal{H}_n}\\
    \hat{X}_\mathcal{E} = \hat{X}_\mathcal{E}^\dagger \succeq 0, \,\operatorname{Tr}_{\mathcal{H}_n}\hat{X}_\mathcal{E} = \hat{I}_{\mathcal{H}_0}.
    \end{align}

\begin{figure}
\centering
\includegraphics[width=0.5\textwidth]{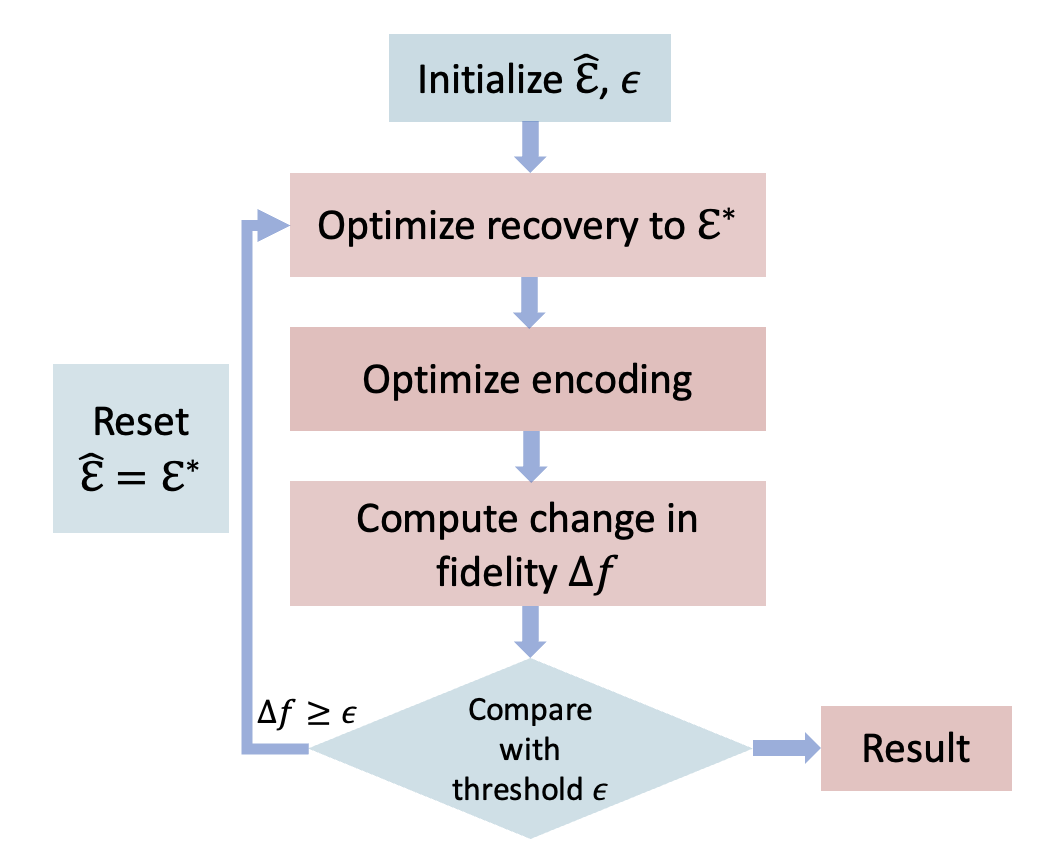}
\caption{\label{fig:0001} flowchart for alternating semidefinite program.}
\end{figure}

While global optimization offers a solution \cite{FloudasVisweswaran1990}, its exponential complexity renders it impractical for this context. As an alternative, one can solve for encoding and decoding in a heuristic manner as illustrated in Figure~\ref{fig:0001}. The process involves iterative optimization of the decoding and encoding maps, starting from a randomly initialized encoding map. The optimization was conducted using semidefinite programming via CVX \cite{CVXResearch2012, GrantBoyd2008}, a tool for convex optimization. The result is a optimized four-qubit code with both optimized encoding and decoding maps.

\section{Four-Qubit Quantum Error Correction Code}
\subsection{\label{sec:level2}Optimized four-qubit codeword}

Using biconvex optimization \cite{Noh2018,KosutLidar2009},  we refine the ansatz for a four-qubit codeword tailored to correct amplitude damping error:
\begin{align}
    |0_L\rangle & \equiv \sqrt{1-\frac{1}{2(1-\gamma)^2}}|0000\rangle + \frac{1}{\sqrt{2}(1-\gamma)}|1111\rangle \\
    |1_L\rangle & \equiv \frac{1}{2}(|0011\rangle + |0101\rangle - |1010\rangle + |1100\rangle).
\end{align}
We verified this expression by numerically comparing with the biconvex optimization solution for a range of $\gamma$ from [$0.01$, $0.1$], as illustrated in Appendix Fig. \ref{fig:SDP}. This expression of codeword works for $\gamma \ll 1$.

\subsection{Approximated QEC Criteria}
The QEC criteria specify conditions for a code to successfully recover quantum states after error occurrence. For a code \( \mathcal{C} \) with error operators \( \mathcal{E} = \{E_a\} \), the criteria are met if:
\begin{equation}
\langle \psi_i | E_a^\dagger E_b | \psi_j \rangle = C_{ab} \delta_{ij},
\end{equation}
for all \( |\psi_i\rangle, |\psi_j\rangle \in \mathcal{C} \), where \( \delta_{ij} \) is the Kronecker delta and \( C_{ab} \) is a constant dependent only on \( E_a \) and \( E_b \).


In the optimized four-qubit code for amplitude damping
errors case, we show that it satisfies the approximate QEC criteria. We prove:
\begin{equation}
\langle \psi_i | E_a^\dagger E_b | \psi_j \rangle = C_{ab} \delta_{ij} + O(\gamma^2),
\label{eq:18}
\end{equation}
for any $E_a, E_b \in \mathcal{E}^{(0)} \cup  \mathcal{E}^{(1)} $. That is, we show that we can suppress the error to second order for no-damping and first-order damping error. Detailed calculations of each $\langle \psi_i | E_a^\dagger E_b | \psi_j \rangle$ are provided in the appendix.

\subsection{Quantum Error Recovery (QER)}
\subsubsection{Optimized QER}
Using the optimized recovery results from convex optimization, we achieve an entanglement fidelity of:
\begin{equation}
    F_{ent} = 1 - 1.09\gamma^2 + O(\gamma^3)
\end{equation} 
which surpasses the \( F_{ent} = 1 - 1.25\gamma^2 + O(\gamma^3) \) achieved by Leung et al.'s four-qubit code with optimized recovery (as illustrated by the solid lines in Fig. \ref{fig:F}).\\

We decompose the optimized recovery and decoding Choi matrix, \( \hat{X}_{\mathcal{D \cdot R}} \), obtained from convex optimization, into Kraus operators. Since \( \hat{X}_{\mathcal{D \cdot R}} \) is positive semi-definite, it can be expressed as:
\begin{equation}
\hat{X}_{\mathcal{D \cdot R}} = \sum_i \lambda_i \, |\phi_i\rangle \langle \phi_i|,
\end{equation}
where \( \lambda_i \) are the eigenvalues, and \( |\phi_i\rangle \) are the eigenvectors.\\

Each eigenvector \( |\phi_i\rangle \) can be reshaped into a Kraus operator. For our \( 16 \times 2 \) quantum decoding channel, which decodes a \( 16\)-dimensional Hilbert space into a \( 2\) -dimensional Hilbert space, we reshape \( |\phi_i\rangle \) into a \( 2 \times 16 \) matrix. Multiplying the reshaped matrix by \( \sqrt{\lambda_i} \) yields the Kraus operator:
\begin{equation}
K_{\mathcal{D \cdot R}, i} = \sqrt{\lambda_i} \, \text{Reshape}(|\phi_i\rangle).
\end{equation}

Due to the constraint
\begin{equation}
\operatorname{Tr}_{\mathcal{H}_0} \hat{X}_{\mathcal{D \cdot R}} = \hat{I}_{\mathcal{H}_n},
\end{equation}the Kraus operators satisfy the trace-preserving condition:
\begin{equation}
\sum_i K_{\mathcal{D \cdot R}, i}^\dagger K_{\mathcal{D \cdot R}, i} = I.
\end{equation}

Finally, we separate the recovery operators from the decoding operators, assuming that the decoding operator is the inverse of the encoding. The recovery operators are not unique. The operators presented here represent only one possible form that satisfies the required conditions.\\

Analyzing the recovery operators corresponding to different values of $\gamma$, we observe that $R_1$ and $R_6$ exhibit a strong dependence on $\gamma$. For the entries of these operators, we fit their $\gamma$-dependence to capture their behavior accurately. By contrast, the other operators show negligible $\gamma$-dependence, allowing us to retain their numerical approximations without modeling $\gamma$-dependence.\\

\noindent\begin{minipage}{0.99\columnwidth}
    \fbox{%
        \parbox{\dimexpr\linewidth-2\fboxsep-2\fboxrule\relax}{
            \tiny
            \begin{align*}
                R_1 &= -\left( |0_L\rangle \left( \frac{\sqrt{1-2\sqrt{2}\,\gamma^2}}{\sqrt{2}} \, \langle 0000 | + \frac{\sqrt{1+2\sqrt{2}\,\gamma^2}}{\sqrt{2}} \, \langle 1111 | \right) \right) - |1_L\rangle \, \langle1_L|
                \\[5pt]
                R_2 &= |0_L\rangle \left( -0.7735 \, \langle 0111 | + 0.0997 \, \langle 1011 | + 0.4077 \, \langle 1101 | - 0.4749 \, \langle 1110 | \right)
                \\ 
                &\quad + |1_L\rangle \left( 0.3588 \, \langle 0001 | + 0.2111 \, \langle 0010 | - 0.8828 \, \langle 0100 | - 0.2177 \, \langle 1000 | \right)
                \\[4pt]
                R_3 &= |0_L\rangle \left( 0.2749 \, \langle 0111 | + 0.6843 \, \langle 1011 | - 0.3314 \, \langle 1101 | - 0.5885 \, \langle 1110 | \right)
                \\ 
                &\quad + |1_L\rangle \left( 0.2495 \, \langle 0001 | - 0.6105 \, \langle 0010 | - 0.2217 \, \langle 0100 | + 0.7182 \, \langle 1000 | \right)
                \\[6pt]
                R_4 &= |0_L\rangle \left( -0.2840 \, \langle 0111 | + 0.7002 \, \langle 1011 | + 0.0506 \, \langle 1101 | + 0.6530 \, \langle 1110 | \right)
                \\ 
                &\quad + |1_L\rangle \left( 0.5309 \, \langle 0001 | + 0.6626 \, \langle 0010 | + 0.2610 \, \langle 0100 | + 0.4594 \, \langle 1000 | \right)
                \\[6pt]
                R_5 &= |0_L\rangle \left( 0.4954 \, \langle 0111 | + 0.1774 \, \langle 1011 | + 0.8493 \, \langle 1101 | - 0.0406 \, \langle 1110 | \right)
                \\ 
                &\quad + |1_L\rangle \left( 0.7260 \, \langle 0001 | - 0.3790 \, \langle 0010 | + 0.3216 \, \langle 0100 | - 0.4752 \, \langle 1000 | \right)
                \\[6pt]
                R_6 &= -\frac{1}{2}|0_L\rangle  \left( \langle 0011 | - \langle 0101 | + \langle 1010 | + \langle 1100 | \right)  
                \\
                &\quad + |1_L\rangle \left( -\frac{\sqrt{1-2\sqrt{2}\,\gamma^2}}{\sqrt{2}} \, \langle 1111 | + \frac{\sqrt{1+2\sqrt{2}\,\gamma^2}}{\sqrt{2}} \, \langle 0000 | \right) 
                \\[6pt]
                R_7 &= |0_L\rangle \left( -0.0823 \left( \langle 0011 | - \langle 1100 | \right) + 0.6412 \left( \langle 0101 | + \langle 1010 | \right) \right.
                \\ 
                &\quad \left. - 0.2866 \left( \langle 0110 | - \langle 1001 | \right) \right)
                \\[6pt]
                R_8 &= |0_L\rangle \left( 0.2401 \left( \langle 0011 | - \langle 1100 | \right) - 0.2454 \left( \langle 0101 | + \langle 1010 | \right) \right.
                \\ 
                &\quad \left. - 0.6180 \left( \langle 0110 | - \langle 1001 | \right) \right)
                \\[6pt]
                R_9 &= |0_L\rangle \left( -0.6600 \left( \langle 0011 | - \langle 1100 | \right) - 0.1693 \left( \langle 0101 | + \langle 1010 | \right) \right.
                \\ 
                &\quad \left. - 0.1891 \left( \langle 0110 | - \langle 1001 | \right) \right)
                \\[6pt]
                R_{10} &= |0_L\rangle \left( 0.0011 \left( \langle 0101 | + \langle 1010 | \right) + 0.7071 \left( \langle 0110 | + \langle 1001 | \right) \right)
            \end{align*}
          }  
        }
    \end{minipage}
\\

\subsubsection{Analytical QER}
We also derive an analytical recovery map for the new codeword, which is described as follows:\\

\noindent\begin{minipage}{0.99\columnwidth}
    \fbox{%
        \parbox{\dimexpr\linewidth-2\fboxsep-2\fboxrule\relax}{
            \small
            \begin{align*}
            R_1 &= |0_L\rangle\langle0111| + \frac{1}{\sqrt{2}} |1_L\rangle(-\langle0010| + \langle0100|)\\
            R_2 &= |0_L\rangle\langle1011| + \frac{1}{\sqrt{2}} |1_L\rangle(\langle0001| + \langle1000|)\\
            R_3 &= |0_L\rangle\langle1101| + \frac{1}{\sqrt{2}} |1_L\rangle(\langle0001| - \langle1000|)\\
            R_4 &= |0_L\rangle\langle1110| + \frac{1}{\sqrt{2}} |1_L\rangle(\langle0010| + \langle0100|)\\
            R_5 &= |0_L\rangle\langle1001| \\
            R_6 &= |0_L\rangle\langle0110| \\
            R_7 &= |0_L\rangle(\alpha\langle0000| + \beta\langle1111|) +  |1_L\rangle\langle1_L|\\
            R_8 &= |0_L\rangle(\beta\langle0000| - \alpha\langle1111|)\\
            &+|\psi_1 \rangle \langle \psi_1|
            +|\psi_2 \rangle \langle \psi_2|
            +|\psi_3 \rangle \langle \psi_3|
            \end{align*}
            \hspace{1cm}{
            \small with
            \begin{align*}
                |\psi_1 \rangle &= \frac{-|0011\rangle + |0101\rangle + |1010\rangle + |1100\rangle}{2}\\
            |\psi_2 \rangle &= \frac{|0011\rangle - |0101\rangle + |1010\rangle + |1100\rangle}{2} \\
            |\psi_3 \rangle &= \frac{|0011\rangle + |0101\rangle + |1010\rangle - |1100\rangle}{2} 
            \end{align*}
          }  
        }
    }
\end{minipage}

\begin{figure}[H]
    \hspace*{-0.4cm}
    \centering
    \includegraphics[width=0.5\textwidth]{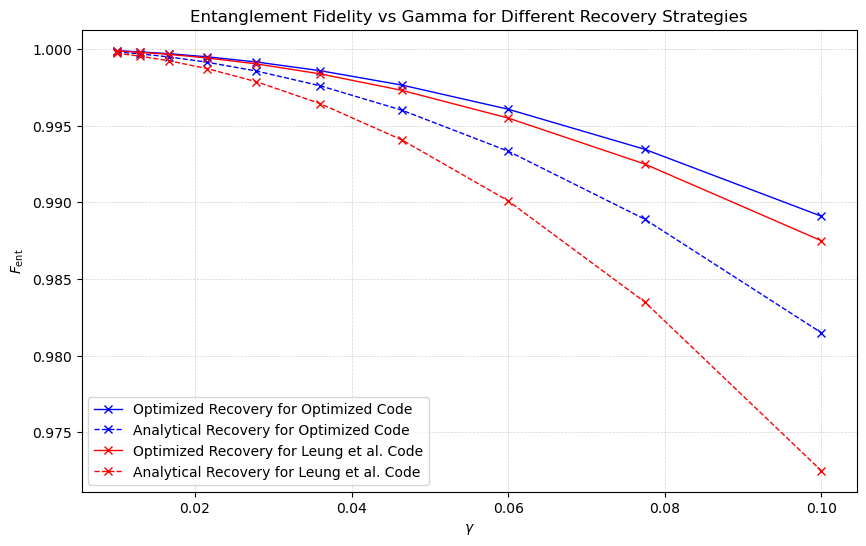}
    \caption{Entanglement Fidelity Comparison Between Optimized Code and Leung et. al's Code. \label{fig:F}}
\end{figure}

Our recovery map $\mathcal{R}(\hat{\rho}) = \sum\limits_{i = 1}^{8}R_i\hat{\rho} R_i^{\dagger}$ is capable of perfectly correcting all first-order damping errors $\mathcal{E}^{(1)}$. Without loss of generality, consider the case where a damping error occurs in the first qubit: 
\begin{equation}
    \mathcal{E}^{(1)}_{1000}|0_L\rangle
    = A_1 \otimes A_0^{\otimes3}|0_L\rangle=\sqrt{\frac{\gamma(1-\gamma)}{2}}|0111\rangle
\label{eq:19}
\end{equation}
\begin{equation}
    \mathcal{E}^{(1)}_{1000}|1_L\rangle=\frac{\sqrt{\gamma(1-\gamma)}}{2}(-|0010\rangle+|0100\rangle).
\label{eq:20}
\end{equation}In this scenario, the recovery operator \(R_1\) effectively projects onto the syndrome subspace corresponding to $\mathcal{E}^{(1)}_{1000}$, restoring the state to the correct logical codewords. For arbitrary initial state $|\Psi\rangle = C_0|0_L\rangle + C_1|1_L\rangle$:
\begin{equation}
    R_1 \mathcal{E}^{(1)}_{1000}|\Psi\rangle \langle \Psi|{\mathcal{E}^{(1)}_{1000}}^{\dagger} R_1^{\dagger} \propto |\Psi\rangle \langle \Psi|.
\end{equation}
Similarly, operators \(R_2\), \(R_3\), and \(R_4\) correct first-order damping errors on the second, third, and fourth qubits, respectively. Operators \(R_5\) and \(R_6\) are designed to address two-qubit damping errors $\mathcal{E}^{(2)}_{0110}$ and $\mathcal{E}^{(2)}_{1001}$ for $|0_L\rangle$. Take $\mathcal{E}^{(2)}_{0110}$ as an example,
\begin{equation}
    \mathcal{E}^{(2)}_{0110}|0_L\rangle = (A_0 \otimes A_1^{\otimes2}\otimes A_0)|0_L\rangle = \frac{\gamma}{\sqrt{2}}|1001\rangle
\end{equation}
\begin{equation}
    R_5 \mathcal{E}^{(2)}_{0110}|0_L\rangle \langle 0_L|{\mathcal{E}^{(2)}_{0110}}^{\dagger} R_5^{\dagger} \propto |0_L\rangle \langle 0_L|.
\end{equation}
Operators \(R_7\) and \(R_8\) are tailored to correct the distortion in the logical codewords when no-jump error occurs. The arbitrary state \(|\Psi\rangle\), when subject to a no-jump error and subsequently recovered, transforms as:
\begin{equation*}
R_7 \mathcal{E}^{(0)}|\Psi\rangle = [(\alpha + \beta)\sqrt{1 - \frac{1}{2(1 - \gamma)^2}} +(\beta - \alpha)\frac{(1 - \gamma)}{\sqrt{2}}]C_0 |0_L\rangle
\end{equation*}
\begin{equation}
    + \frac{(1 - \gamma)}{2}C_1|1_L\rangle.
\end{equation}
To correct the unbalanced coefficient caused by the no-jump error between $|0_L\rangle$ and $|1_L\rangle$, we apply $R_7$. While it's impossible to completely remove the discrepancy, we optimize \(\alpha\) and \(\beta\) to minimize it, resulting in:
\begin{equation}
\min_{\alpha, \beta} = |(\alpha + \beta)\sqrt{1 - \frac{1}{2(1 - \gamma)^2}} +(\beta - \alpha)\frac{(1 - \gamma)}{\sqrt{2}} -\frac{(1 - \gamma)}{\sqrt{2}}|
\end{equation}
where \(\alpha^2 + \beta^2 = 1\).

\begin{equation}
    \alpha \approx \sqrt{1-\frac{1}{2(1-\gamma)^2}} + 0.71\gamma + 0.76\gamma^2 + O(\gamma^3),
\end{equation}
\begin{equation}
    \beta = \sqrt{1-\alpha^2}.
\end{equation}
This analytical QER approach enables our optimized code to achieve an entanglement fidelity of \(F_{ent} = 1 - 1.85\gamma^2 + O(\gamma^3)\), significantly surpassing the \(F_{ent} = 1 - 2.75\gamma^2 + O(\gamma^3)\) of Leung et al.'s analytically recovered code, as the dashed lines in Fig. \ref{fig:F}.

\section{Conclusion}

In this study, we have presented a novel four-qubit quantum error correcting code optimized for the amplitude damping channel, a prevalent source of noise in quantum systems such as superconducting circuits. Our approach, grounded in biconvex optimization, leverages the technique of alternating semi-definite programming to achieve a high-performance, noise-adapted QEC scheme.

The optimized code demonstrates a significant improvement in entanglement fidelity over the Leung-Nielsen-Chuang-Yamamoto four-qubit code. The capability of our code to efficiently handle both one- and two-qubit damping errors up to $O(\gamma)$ underscores its robustness and suitability for quantum systems where energy dissipation is a major concern.

Furthermore, the analytical recovery channels we have constructed, which closely align with their numerically optimized counterparts, provide a solid foundation for practical implementation. These channels facilitate precise error correction, enhancing the reliability of quantum information processing in environments characterized by amplitude damping.

In conclusion, our work not only contributes to the advancement of quantum error correction in theory but also has implications for the practical realization of reliable quantum computing, especially in the NISQ era. The methods and results discussed herein pave the way for further research and development in quantum error correction, particularly for quantum systems where structured errors like amplitude damping are predominant.

\begin{acknowledgments}
We acknowledge support from the ARO(W911NF-23-1-0077), ARO MURI (W911NF-21-1-0325), AFOSR MURI (FA9550-19-1-0399, FA9550-21-1-0209, FA9550-23-1-0338), DARPA (HR0011-24-9-0359, HR0011-24-9-0361), NSF (OMA-1936118, ERC-1941583, OMA-2137642, OSI-2326767, CCF-2312755), NTT Research, Samsung GRO, Packard Foundation (2020-71479). We are also grateful for the support of the University of Chicago Research Computing Center for assistance with the numerical simulations carried out in this paper. And we acknowledge the fruitful discussion with Ming Yuan, Jerry Zheng, and Pei Zeng. 

\end{acknowledgments}

\cleardoublepage
\appendix*

\section{Derivation of Approximate QEC Criteria}

In this section, we first consider the general error set $\mathcal{E} = \mathcal{E}^{(0)} \cup \mathcal{E}^{(1)} \cup \mathcal{E}^{(2)}$. We derive: 
\begin{align}
\label{eq:A1}
\langle 0_L | E_a^\dagger E_b | 1_L \rangle &= O(\gamma) \\
\langle 0_L | E_a^\dagger E_b | 0_L \rangle &= \langle 1_L | E_a^\dagger E_b | 1_L \rangle + O(\gamma^2)
\label{eq:A2}
\end{align}
Thus, we can show:
\begin{equation}
\langle \psi_i | E_a^\dagger E_b | \psi_j \rangle = C_{ab} \delta_{ij} + O(\gamma)
\label{eq:A3}
\end{equation}
for any $E_a, E_b \in \mathcal{E}$. To do this, we analyze $11 \times 11$ matrices for $\langle 0_L | E_a^\dagger E_b | 1_L \rangle$, $\langle 0_L | E_a^\dagger E_b | 0_L \rangle$, and $\langle 1_L | E_a^\dagger E_b | 1_L \rangle$. These matrices represent the interactions between logical qubit states under $11$ possible error cases, denoted as $\mathcal{E} = \mathcal{E}^{(0)} \cup \mathcal{E}^{(1)} \cup \mathcal{E}^{(2)} = $
$\{\mathcal{E}^{(0)}_{0000}, \mathcal{E}^{(1)}_{1000},\mathcal{E}^{(1)}_{0100},\mathcal{E}^{(1)}_{0010},\mathcal{E}^{(1)}_{0001},\\\mathcal{E}^{(2)}_{1100},\mathcal{E}^{(2)}_{1010},\mathcal{E}^{(2)}_{1001},\mathcal{E}^{(2)}_{0110},\mathcal{E}^{(2)}_{0101}, \mathcal{E}^{(2)}_{0011}\}$. In the matrices, rows and columns correspond to $E_a$ and $E_b$ choices from $\mathcal{E}$. Specifically, the matrix's first row and column represent the no-damping error, rows and columns two through five correspond to first-order damping errors, and rows and columns six to eleven correspond to second-order damping errors. \\

For the matrix $\langle 0_L | E_a^\dagger E_b | 1_L \rangle$, the non-zero entries are defined as:
\begin{equation}
\left\{
\begin{aligned}
    &A_{1,6} = A_{1,7} = A_{1,11} = \frac{1}{2\sqrt{2}}\gamma (1 - \gamma)^2, \\
    &A_{1,10} = -\frac{1}{2\sqrt{2}}\gamma (1 - \gamma)^2, \\
    &A_{6,1} = A_{10,1} = A_{11,1} = \frac{\gamma (1 - \gamma)}{2} \sqrt{1 - \frac{1}{2 (1 - \gamma)^{2}}}, \\
    &A_{7,1} = -\frac{\gamma (1 - \gamma)}{2} \sqrt{1 - \frac{1}{2 (1 - \gamma)^{2}}}.
    \label{eq:A4}
\end{aligned}
\right.
\end{equation}
All other entries of this matrix are zero,
with the non-zero terms being of order at most $O(\gamma)$. Hence, Eq.(~\ref{eq:A1}) is satisfied for all $E_a, E_b \in \mathcal{E}$.\\ 

For the matrix $\langle 0_L | E_a^\dagger E_b | 0_L \rangle$, the non-zero entries are confined to the diagonal and detailed as follows: 
\begin{equation}
\left\{
    \begin{aligned}
        &A_{i,i} = 
        \begin{cases}
            1 + \frac{1}{2} (1-\gamma)^2 - \frac{1}{2 (1 - \gamma)^{2}}\\
            \approx 1 - 2\gamma + O(\gamma^2), & i = 1\\
            \gamma(1 - \frac{1}{2 (1 - \gamma)^{2}})\\
            \approx \frac{\gamma}{2} - \gamma^2 + O(\gamma^3), & 2 \leq i \leq 5\\
            \gamma^2(1 - \frac{1}{2 (1 - \gamma)^{2}})\\
            \approx \frac{\gamma^2}{2} + O(\gamma^3), & 6 \leq i \leq 11
        \end{cases} \\
        &A_{i,j} = 0. & \hspace{-100mm} i \neq j
    \end{aligned}
\right.
\end{equation}

For the matrix $\langle 1_L | E_a^\dagger E_b | 1_L \rangle$, the non-zero entries are as follows:
\begin{equation}
\left\{
    \begin{aligned}
        &A_{i,i} = 
        \begin{cases}
            1 - 2\gamma +\gamma^2, & i = 1\\
            \frac{\gamma}{2} - \gamma^{2}+\frac{\gamma^3}{2}, & 2 \leq i \leq 5\\
            \frac{\gamma^{2}}{4} - \frac{\gamma^3}{2}+ \frac{\gamma^4}{4}, & i = 6, 7, 10, 11
        \end{cases} \\
        &A_{6,7} = A_{6,11} = A_{7,6} = A_{7,11} = A_{11,6} = A_{11,7} \\
        &= \frac{\gamma^{2}}{4} - \frac{\gamma^3}{2}+ \frac{\gamma^4}{4}, \\
        &A_{6,10} = A_{7,10} = A_{10,6} = A_{10,7} = A_{10,11} = A_{11,10} \\
        &= -(\frac{\gamma^{2}}{4} - \frac{\gamma^3}{2}+ \frac{\gamma^4}{4}).
    \end{aligned}
\right.
\end{equation}

From the calculations of $\langle 0_L | E_a^\dagger E_b | 0_L \rangle$ and $\langle 1_L | E_a^\dagger E_b | 1_L \rangle$, we find that $\langle 0_L | E_a^\dagger E_b | 0_L \rangle = \langle 1_L | E_a^\dagger E_b | 1_L \rangle + O(\gamma^2)$. This result fulfills the requirement of Eq.(~\ref{eq:A2}).\\

To derive Eq.(~\ref{eq:18}), we focus on the subset $E_a, E_b \in \mathcal{E}^{(0)}\cup \mathcal{E}^{(1)}$, corresponding to the initial five rows and columns within the $11 \times 11$ matrix. From Eq.(~\ref{eq:A4}), it follows that:
\begin{equation}
\langle 0_L | E_a^\dagger E_b | 1_L \rangle = 0, \quad E_a, E_b \in \mathcal{E}^{(0)}\cup \mathcal{E}^{(1)}
\end{equation}
This, in conjunction with Eq.(~\ref{eq:A2}), yields Eq.(~\ref{eq:18}).\\

Let $M$ denote the QEC Matrix and $\Delta M$ denote its deviation from satisfying perfect QEC criteria. We find that: 
\begin{equation}
   \max\limits_{i,j}|\Delta M _{i,j}| = \frac{1}{2\sqrt{2}}\gamma + O(\gamma^2) 
\end{equation}
indicating a smaller magnitude of deviation from perfect QEC criteria compared to the code by Leung et al\cite{Leung1997}. For their code, the deviation is given by:
\begin{equation}
   \max\limits_{i,j}|\Delta M _{i,j}| = \frac{1}{2}\gamma + O(\gamma^2) 
\end{equation}
Ref.\cite{zheng2024nearoptimal} elucidates that evaluating \(\Delta M\) can provide a narrow two-sided bound for optimal code fidelity. Consequently, the reduced deviation observed in our code is heuristically linked to enhanced fidelity over Leung et al.'s code.

\begin{figure}[H]
    \hspace*{-0.4cm}
    \centering
    \includegraphics[width=0.5\textwidth]{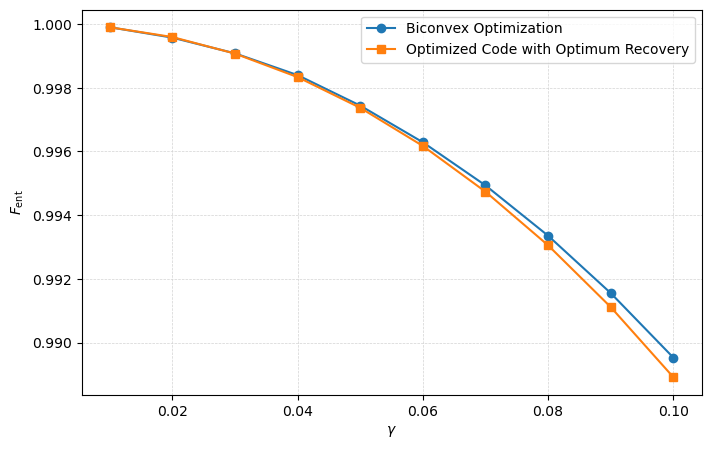}
    \caption{Comparison between Optimized Four-qubit Codeword and Biconvex Optimization Solution \label{fig:SDP}}
\end{figure}

\newpage
\nocite{*}


\begin{thebibliography}{21}%
\makeatletter
\providecommand \@ifxundefined [1]{%
 \@ifx{#1\undefined}
}%
\providecommand \@ifnum [1]{%
 \ifnum #1\expandafter \@firstoftwo
 \else \expandafter \@secondoftwo
 \fi
}%
\providecommand \@ifx [1]{%
 \ifx #1\expandafter \@firstoftwo
 \else \expandafter \@secondoftwo
 \fi
}%
\providecommand \natexlab [1]{#1}%
\providecommand \enquote  [1]{``#1''}%
\providecommand \bibnamefont  [1]{#1}%
\providecommand \bibfnamefont [1]{#1}%
\providecommand \citenamefont [1]{#1}%
\providecommand \href@noop [0]{\@secondoftwo}%
\providecommand \href [0]{\begingroup \@sanitize@url \@href}%
\providecommand \@href[1]{\@@startlink{#1}\@@href}%
\providecommand \@@href[1]{\endgroup#1\@@endlink}%
\providecommand \@sanitize@url [0]{\catcode `\\12\catcode `\$12\catcode `\&12\catcode `\#12\catcode `\^12\catcode `\_12\catcode `\%12\relax}%
\providecommand \@@startlink[1]{}%
\providecommand \@@endlink[0]{}%
\providecommand \url  [0]{\begingroup\@sanitize@url \@url }%
\providecommand \@url [1]{\endgroup\@href {#1}{\urlprefix }}%
\providecommand \urlprefix  [0]{URL }%
\providecommand \Eprint [0]{\href }%
\providecommand \doibase [0]{http://dx.doi.org/}%
\providecommand \selectlanguage [0]{\@gobble}%
\providecommand \bibinfo  [0]{\@secondoftwo}%
\providecommand \bibfield  [0]{\@secondoftwo}%
\providecommand \translation [1]{[#1]}%
\providecommand \BibitemOpen [0]{}%
\providecommand \bibitemStop [0]{}%
\providecommand \bibitemNoStop [0]{.\EOS\space}%
\providecommand \EOS [0]{\spacefactor3000\relax}%
\providecommand \BibitemShut  [1]{\csname bibitem#1\endcsname}%
\let\auto@bib@innerbib\@empty
\bibitem [{\citenamefont {Bennett}\ \emph {et~al.}(1996)\citenamefont {Bennett}, \citenamefont {DiVincenzo}, \citenamefont {Smolin},\ and\ \citenamefont {Wootters}}]{BennettEtAl1996}%
  \BibitemOpen
  \bibfield  {author} {\bibinfo {author} {\bibfnamefont {C.~H.}\ \bibnamefont {Bennett}}, \bibinfo {author} {\bibfnamefont {D.~P.}\ \bibnamefont {DiVincenzo}}, \bibinfo {author} {\bibfnamefont {J.~A.}\ \bibnamefont {Smolin}}, \ and\ \bibinfo {author} {\bibfnamefont {W.~K.}\ \bibnamefont {Wootters}},\ }\href@noop {} {\ \textbf {\bibinfo {volume} {54}} (\bibinfo {year} {1996})},\ \bibinfo {note} {lANL e-print quant-ph/9604024}\BibitemShut {NoStop}%
\bibitem [{\citenamefont {Ekert}\ and\ \citenamefont {Macchiavello}(1996)}]{EkertMacchiavello1996}%
  \BibitemOpen
  \bibfield  {author} {\bibinfo {author} {\bibfnamefont {A.}~\bibnamefont {Ekert}}\ and\ \bibinfo {author} {\bibfnamefont {C.}~\bibnamefont {Macchiavello}},\ }\href@noop {} {\bibfield  {journal} {\bibinfo  {journal} {Physical Review Letters}\ }\textbf {\bibinfo {volume} {77}},\ \bibinfo {pages} {2585} (\bibinfo {year} {1996})}\BibitemShut {NoStop}%
\bibitem [{\citenamefont {Knill}\ and\ \citenamefont {Laflamme}(1996)}]{KnillLaflamme1996}%
  \BibitemOpen
  \bibfield  {author} {\bibinfo {author} {\bibfnamefont {E.}~\bibnamefont {Knill}}\ and\ \bibinfo {author} {\bibfnamefont {R.}~\bibnamefont {Laflamme}},\ }\href@noop {} {\bibfield  {journal} {\bibinfo  {journal} {Physical Review A}\ } (\bibinfo {year} {1996})},\ \bibinfo {note} {lANL E-print quant-ph/9604034}\BibitemShut {NoStop}%
\bibitem [{\citenamefont {Nielsen}\ \emph {et~al.}(1996)\citenamefont {Nielsen}, \citenamefont {Barnum}, \citenamefont {Caves},\ and\ \citenamefont {Schumacher}}]{NielsenEtAl1996}%
  \BibitemOpen
  \bibfield  {author} {\bibinfo {author} {\bibfnamefont {M.~A.}\ \bibnamefont {Nielsen}}, \bibinfo {author} {\bibfnamefont {H.}~\bibnamefont {Barnum}}, \bibinfo {author} {\bibfnamefont {C.~M.}\ \bibnamefont {Caves}}, \ and\ \bibinfo {author} {\bibfnamefont {B.~W.}\ \bibnamefont {Schumacher}},\ }\href@noop {} {} (\bibinfo {year} {1996}),\ \bibinfo {note} {in preparation}\BibitemShut {NoStop}%
\bibitem [{\citenamefont {Calderbank}\ \emph {et~al.}(1996)\citenamefont {Calderbank}, \citenamefont {Rains}, \citenamefont {Shor},\ and\ \citenamefont {Sloane}}]{CalderbankEtAl1996}%
  \BibitemOpen
  \bibfield  {author} {\bibinfo {author} {\bibfnamefont {A.~R.}\ \bibnamefont {Calderbank}}, \bibinfo {author} {\bibfnamefont {E.~M.}\ \bibnamefont {Rains}}, \bibinfo {author} {\bibfnamefont {P.~W.}\ \bibnamefont {Shor}}, \ and\ \bibinfo {author} {\bibfnamefont {N.~J.~A.}\ \bibnamefont {Sloane}},\ }\href@noop {} {} (\bibinfo {year} {1996}),\ \bibinfo {note} {lANL E-print quant-ph/9608006}\BibitemShut {NoStop}%
\bibitem [{\citenamefont {Schumacher}(1996)}]{Schumacher1996}%
  \BibitemOpen
  \bibfield  {author} {\bibinfo {author} {\bibfnamefont {B.}~\bibnamefont {Schumacher}},\ }\href {\doibase 10.1103/PhysRevA.54.2614} {\bibfield  {journal} {\bibinfo  {journal} {Phys. Rev. A}\ }\textbf {\bibinfo {volume} {54}},\ \bibinfo {pages} {2614} (\bibinfo {year} {1996})}\BibitemShut {NoStop}%
\bibitem [{\citenamefont {Leung}\ \emph {et~al.}(1997)\citenamefont {Leung}, \citenamefont {Nielsen}, \citenamefont {Chuang},\ and\ \citenamefont {Yamamoto}}]{Leung1997}%
  \BibitemOpen
  \bibfield  {author} {\bibinfo {author} {\bibfnamefont {D.~W.}\ \bibnamefont {Leung}}, \bibinfo {author} {\bibfnamefont {M.~A.}\ \bibnamefont {Nielsen}}, \bibinfo {author} {\bibfnamefont {I.~L.}\ \bibnamefont {Chuang}}, \ and\ \bibinfo {author} {\bibfnamefont {Y.}~\bibnamefont {Yamamoto}},\ }\href {\doibase 10.1103/PhysRevA.56.2567} {\bibfield  {journal} {\bibinfo  {journal} {Physical Review A}\ }\textbf {\bibinfo {volume} {56}},\ \bibinfo {pages} {2567} (\bibinfo {year} {1997})}\BibitemShut {NoStop}%
\bibitem {DuttaMandayam2025}%
  \BibitemOpen
  \bibfield  {author} {\bibinfo {author} {\bibfnamefont {S.}~\bibnamefont {Dutta}}, \bibinfo {author} {\bibfnamefont {P.}~\bibnamefont {Mandayam}},\ and\ \bibinfo {author} {\bibfnamefont {A.}~\bibnamefont {Jain}},\ }\href {https://arxiv.org/abs/2410.00155} {\ \bibinfo {title} {Smallest quantum codes for amplitude-damping noise}},\ \bibinfo {note} {arXiv:2410.00155 [quant-ph] (2024)}\BibitemShut {NoStop}%
\bibitem [{\citenamefont {Fletcher}\ \emph {et~al.}(2007{\natexlab{a}})\citenamefont {Fletcher}, \citenamefont {Shor},\ and\ \citenamefont {Win}}]{Fletcher2007}%
  \BibitemOpen
  \bibfield  {author} {\bibinfo {author} {\bibfnamefont {A.~S.}\ \bibnamefont {Fletcher}}, \bibinfo {author} {\bibfnamefont {P.~W.}\ \bibnamefont {Shor}}, \ and\ \bibinfo {author} {\bibfnamefont {M.~Z.}\ \bibnamefont {Win}},\ }\href {\doibase 10.1103/PhysRevA.75.012338} {\bibfield  {journal} {\bibinfo  {journal} {Physical Review A}\ }\textbf {\bibinfo {volume} {75}},\ \bibinfo {pages} {012338} (\bibinfo {year} {2007}{\natexlab{a}})}\BibitemShut {NoStop}%
\bibitem [{\citenamefont {Noh}\ \emph {et~al.}(2018)\citenamefont {Noh}, \citenamefont {Albert},\ and\ \citenamefont {Jiang}}]{Noh2018}%
  \BibitemOpen
  \bibfield  {author} {\bibinfo {author} {\bibfnamefont {K.}~\bibnamefont {Noh}}, \bibinfo {author} {\bibfnamefont {V.~V.}\ \bibnamefont {Albert}}, \ and\ \bibinfo {author} {\bibfnamefont {L.}~\bibnamefont {Jiang}},\ }\href {https://ar5iv.org/pdf/1801.07271} {\bibfield  {journal} {\bibinfo  {journal} {arXiv}\ } (\bibinfo {year} {2018})},\ \Eprint {http://arxiv.org/abs/1801.07271} {1801.07271} \BibitemShut {NoStop}%
\bibitem [{\citenamefont {Kosut}\ and\ \citenamefont {Lidar}(2009)}]{KosutLidar2009}%
  \BibitemOpen
  \bibfield  {author} {\bibinfo {author} {\bibfnamefont {R.~L.}\ \bibnamefont {Kosut}}\ and\ \bibinfo {author} {\bibfnamefont {D.~A.}\ \bibnamefont {Lidar}},\ }\href@noop {} {\bibfield  {journal} {\bibinfo  {journal} {Quantum Information Processing}\ }\textbf {\bibinfo {volume} {8}},\ \bibinfo {pages} {443} (\bibinfo {year} {2009})}\BibitemShut {NoStop}%
\bibitem [{\citenamefont {Reimpell}\ and\ \citenamefont {Werner}(2005)}]{ReimpellWerner2005}%
  \BibitemOpen
  \bibfield  {author} {\bibinfo {author} {\bibfnamefont {M.}~\bibnamefont {Reimpell}}\ and\ \bibinfo {author} {\bibfnamefont {R.~F.}\ \bibnamefont {Werner}},\ }\href@noop {} {\bibfield  {journal} {\bibinfo  {journal} {Physical Review Letters}\ }\textbf {\bibinfo {volume} {94}},\ \bibinfo {pages} {080501} (\bibinfo {year} {2005})}\BibitemShut {NoStop}%
\bibitem [{\citenamefont {Choi}(1975)}]{Choi1975}%
  \BibitemOpen
  \bibfield  {author} {\bibinfo {author} {\bibfnamefont {M.-D.}\ \bibnamefont {Choi}},\ }\href@noop {} {\bibfield  {journal} {\bibinfo  {journal} {Linear Algebra and its Applications}\ }\textbf {\bibinfo {volume} {10}},\ \bibinfo {pages} {285} (\bibinfo {year} {1975})}\BibitemShut {NoStop}%
\bibitem [{\citenamefont {Floudas}\ and\ \citenamefont {Visweswaran}(1990)}]{FloudasVisweswaran1990}%
  \BibitemOpen
  \bibfield  {author} {\bibinfo {author} {\bibfnamefont {C.~A.}\ \bibnamefont {Floudas}}\ and\ \bibinfo {author} {\bibfnamefont {V.}~\bibnamefont {Visweswaran}},\ }\href@noop {} {\bibfield  {journal} {\bibinfo  {journal} {Computers \& Chemical Engineering}\ }\textbf {\bibinfo {volume} {14}},\ \bibinfo {pages} {1397} (\bibinfo {year} {1990})}\BibitemShut {NoStop}%
\bibitem [{\citenamefont {{CVX Research, Inc.}}(2012)}]{CVXResearch2012}%
  \BibitemOpen
  \bibfield  {author} {\bibinfo {author} {\bibnamefont {{CVX Research, Inc.}}},\ }\href {URL where the software can be accessed} {\enquote {\bibinfo {title} {Cvx: Matlab software for disciplined convex programming, version 2.0},}\ } (\bibinfo {year} {2012}),\ \bibinfo {note} {[Online; accessed 20-Jan-2024]}\BibitemShut {NoStop}%
\bibitem [{\citenamefont {Grant}\ and\ \citenamefont {Boyd}(2008)}]{GrantBoyd2008}%
  \BibitemOpen
  \bibfield  {author} {\bibinfo {author} {\bibfnamefont {M.}~\bibnamefont {Grant}}\ and\ \bibinfo {author} {\bibfnamefont {S.}~\bibnamefont {Boyd}},\ }in\ \href@noop {} {\emph {\bibinfo {booktitle} {Recent Advances in Learning and Control}}},\ \bibinfo {editor} {edited by\ \bibinfo {editor} {\bibfnamefont {V.}~\bibnamefont {Blondel}}, \bibinfo {editor} {\bibfnamefont {S.}~\bibnamefont {Boyd}}, \ and\ \bibinfo {editor} {\bibfnamefont {H.}~\bibnamefont {Kimura}}}\ (\bibinfo  {publisher} {Springer-Verlag Limited},\ \bibinfo {year} {2008})\ pp.\ \bibinfo {pages} {95--110}\BibitemShut {NoStop}%
\bibitem [{\citenamefont {Zheng}\ \emph {et~al.}(2024)\citenamefont {Zheng}, \citenamefont {He}, \citenamefont {Lee},\ and\ \citenamefont {Jiang}}]{zheng2024nearoptimal}%
  \BibitemOpen
  \bibfield  {author} {\bibinfo {author} {\bibfnamefont {G.}~\bibnamefont {Zheng}}, \bibinfo {author} {\bibfnamefont {W.}~\bibnamefont {He}}, \bibinfo {author} {\bibfnamefont {G.}~\bibnamefont {Lee}}, \ and\ \bibinfo {author} {\bibfnamefont {L.}~\bibnamefont {Jiang}},\ }\href@noop {} {\enquote {\bibinfo {title} {The near-optimal performance of quantum error correction codes},}\ } (\bibinfo {year} {2024}),\ \Eprint {http://arxiv.org/abs/2401.02022} {arXiv:2401.02022 [quant-ph]} \BibitemShut {NoStop}%
\bibitem [{\citenamefont {Yamamoto}\ \emph {et~al.}(2005)\citenamefont {Yamamoto}, \citenamefont {Hara},\ and\ \citenamefont {Tsumura}}]{Yamamoto2005}%
  \BibitemOpen
  \bibfield  {author} {\bibinfo {author} {\bibfnamefont {N.}~\bibnamefont {Yamamoto}}, \bibinfo {author} {\bibfnamefont {S.}~\bibnamefont {Hara}}, \ and\ \bibinfo {author} {\bibfnamefont {K.}~\bibnamefont {Tsumura}},\ }\href@noop {} {\bibfield  {journal} {\bibinfo  {journal} {Physical Review A}\ }\textbf {\bibinfo {volume} {71}},\ \bibinfo {pages} {022322} (\bibinfo {year} {2005})}\BibitemShut {NoStop}%
\bibitem [{\citenamefont {Audenaert}\ and\ \citenamefont {De~Moor}(2002)}]{Audenaert2002}%
  \BibitemOpen
  \bibfield  {author} {\bibinfo {author} {\bibfnamefont {K.}~\bibnamefont {Audenaert}}\ and\ \bibinfo {author} {\bibfnamefont {B.}~\bibnamefont {De~Moor}},\ }\href@noop {} {\bibfield  {journal} {\bibinfo  {journal} {Physical Review A}\ }\textbf {\bibinfo {volume} {65}},\ \bibinfo {pages} {030302} (\bibinfo {year} {2002})},\ \bibinfo {note} {rapid Communications}\BibitemShut {NoStop}%
\bibitem [{\citenamefont {Preskill}(2005)}]{Preskill2005}%
  \BibitemOpen
  \bibfield  {author} {\bibinfo {author} {\bibfnamefont {J.}~\bibnamefont {Preskill}},\ }\href {URL where the notes can be accessed} {\enquote {\bibinfo {title} {Lecture note for quantum computation, chapter 3. foundations of quantum theory ii: Measurement and evolution},}\ } (\bibinfo {year} {2005}),\ \bibinfo {note} {[Online; accessed 20-Jan-2024]}\BibitemShut {NoStop}%
\bibitem [{\citenamefont {Fletcher}\ \emph {et~al.}(2007{\natexlab{b}})\citenamefont {Fletcher}, \citenamefont {Shor},\ and\ \citenamefont {Win}}]{FletcherShorWin2007}%
  \BibitemOpen
  \bibfield  {author} {\bibinfo {author} {\bibfnamefont {A.~S.}\ \bibnamefont {Fletcher}}, \bibinfo {author} {\bibfnamefont {P.~W.}\ \bibnamefont {Shor}}, \ and\ \bibinfo {author} {\bibfnamefont {M.~Z.}\ \bibnamefont {Win}},\ }\href@noop {} {\bibfield  {journal} {\bibinfo  {journal} {Physical Review A}\ }\textbf {\bibinfo {volume} {75}},\ \bibinfo {pages} {012338} (\bibinfo {year} {2007}{\natexlab{b}})}\BibitemShut {NoStop}%
\bibitem [{\citenamefont {Preskill}(2018)}]{Preskill2018}%
  \BibitemOpen
  \bibfield  {author} {\bibinfo {author} {\bibfnamefont {J.}~\bibnamefont {Preskill}},\ }\href {https://ar5iv.org/abs/1801.00862} {\  (\bibinfo {year} {2018})},\ \bibinfo {note} {based on a Keynote Address at Quantum Computing for Business, 5 December 2017}\BibitemShut {NoStop}%
  
\end{thebibliography}
\providecommand{\noopsort}[1]{}\providecommand{\singleletter}[1]{#1}%

\end{document}